\documentclass{llncs}
\usepackage[cmex10]{amsmath}
\usepackage{amssymb}
\usepackage{graphicx}
\usepackage{amssymb,amsmath}
\usepackage{cite}
\usepackage{multirow}

\newcommand{\bldH}{{\mbox{\boldmath $H$}}}

\newcommand{\bldB}{{\mbox{\boldmath $B$}}}
\newcommand{\bldP}{{\mbox{\boldmath $P$}}}
\usepackage{tikz}

\def\remark{
  \let\go\relax
  \ifvmode\vskip-\lastskip\fi
  \noindent{\it Remark\/.}%
  \enskip\relax\ignorespaces\go}
\newcommand{\bs}[1]{\ensuremath{\boldsymbol{#1}}}

\begin{document}

\title{Distance Properties of  Short  LDPC Codes and their Impact on the BP, ML and Near-ML 
Decoding Performance}
\titlerunning{Distance Properties of  Short  LDPC Codes}
\author{
Irina E. Bocharova\inst{1,2} \and 
 Boris D. Kudryashov\inst{1} \and Vitaly Skachek \inst{2} \and
Yauhen Yakimenka\inst{2}}

\institute{St. Petersburg University of Information Technologies,\\
		Mechanics and Optics                                \\
		St. Petersburg 197101, Russia  \\                 
\email{irinaboc@ut.ee, kudryashov\_boris@bk.ru}\\
\and
University of Tartu, Tartu 50409, Estonia\\
\email{vitaly@ut.ee, yauhen@ut.ee}}


\maketitle


\begin{abstract}
Parameters of LDPC codes, such as minimum distance, stopping distance,  stopping 
redundancy, girth of the Tanner graph, and their influence on the frame error rate performance of 
the BP, ML and near-ML decoding over a BEC and an AWGN channel are studied. Both random and structured LDPC codes
are considered. In particular, the BP decoding is applied to the code parity-check matrices with an increasing number of redundant rows, and the convergence of the performance to that of the ML decoding is analyzed. A comparison of the simulated BP, ML, and near-ML performance with the improved theoretical bounds on the error probability based on the exact weight spectrum coefficients and the exact stopping size spectrum coefficients is presented. It is observed that decoding performance very close to the ML decoding performance can be achieved with a relatively small number of redundant rows for some codes, for both the BEC and the AWGN channels.  
 
\end{abstract}

\section{Introduction }
\let\thefootnote\relax\footnote{This work is supported in part by the Norwegian-Estonian Research Cooperation
Programme under the grant EMP133 and by the Estonian Research
Council under the grant PUT405.
} 


 
It is well-known that typically binary LDPC codes have minimum distances which are smaller than those of the best known linear codes of the same rate and length. It is not surprising, since minimum distance does not play an important role in iterative (belief propagation (BP)) decoding. On the other hand, a significant gap in the frame error rate (FER) performance of BP and maximum-likelihood (ML) decoding motivates developing near-ML decoding algorithms for LDPC codes. 

There are  two main approaches  to  improving the BP decoding  performance. First one is based on post-processing in case of  BP decoder failure. Different post-processing techniques  for decoding  of binary LDPC codes over additive white Gaussian noise (AWGN) channels are studied in \cite{pishro2007results,varnica,fang2010bp,bocharova2016low}.  
A similar approach to
decoding of nonbinary LDPC codes over extensions of the binary Galois field is considered in~\cite{baldi2014hybrid}.   
Near-ML decoding algorithms for LDPC codes over binary erasure channel (BEC)  can be found in   
 \cite{pishro2004decoding,hosoya2004,olmos2010tree}. 

The second approach is based on identifying and destroying  specific structural configurations such as trapping and stopping sets of the Tanner graph of the code. In particular, this can be done by adding redundant rows to the code parity-check matrix (see, for example,  \cite{laendner2006cth02,  hehn2010improved, mu2011}).   

Suboptimality of the above modifications of BP decoding rises the following question: which properties of LDPC codes and to what extent influence their decoding FER performance? In this paper, we are trying to partially answer  this question by studying short LDPC codes.  We consider both binary LDPC codes and binary images of nonbinary 
random LDPC codes over extensions of the binary field,  as well as quasi-cyclic (QC) LDPC codes constructed by using  an  optimization technique in~\cite{Boch2016}. Parameters such as minimum distance, stopping distance,  girth of the Tanner graph and estimates on the stopping redundancy  are tabulated. Near-ML decoding based on adding redundant rows to the code parity-check matrix is analyzed. Simulated over the BEC and the AWGN channel, the FER performance of the BP, ML and  near-ML decoding of these classes of LDPC codes is presented and compared to the improved upper bounds on the performance of ML decoding of regular LDPC codes  
 \cite{bocharova2017performance} over the corresponding channels. 
The presented error probability bounds rely on precise 
average enumerators for given ensembles which makes 
these bounds tighter than known bounds (see e.g. \cite {sason2000}).   
  By using an approach similar to that in \cite{bocharova2017performance}, an improved upper bound on the performance of the BP decoding of binary images of nonbinary regular LDPC codes over BEC  is presented.

The paper  is organized as follows. In Section \ref{prelim}, all necessary notations and definitions are given.  In Section \ref{redundancy}, a near-ML decoding  method, which is based on adding redundant rows to the code parity-check matrix, is revisited.  In Section \ref{bounds}, a recurrent procedure for computing the exact coefficients of the weight and stopping set size spectra is described. The improved upper bound on the ensemble average performance of the BP decoding over BEC is derived. Tables of the computed code parameters along with the simulation results for the BP, ML and near-ML decoding are presented in Section \ref{simulations}.  A comparison with the theoretical bounds is done and conclusions are drawn in Section \ref{discussion}.


\section{Preliminaries}
\label{prelim}
\subsection{Ensembles of binary and binary images of nonbinary regular  LDPC codes}

For a binary linear $[n,k]$ code $\mathcal C$  of rate $R=k/n$ denote by  $r=n-k$ its redundancy.  We use a notation  
 $\{A_{n,w} \}_{0\le w\le n}$ for a set of code  weight enumerators, where $A_{n,w}$ is a number of codewords of  weight $w$. Let $\bldH$ be  an $r\times n$ parity-check matrix which defines $\mathcal C$.

 By viewing $\bldH$ as a biadjacency matrix \cite{biadj},  we obtain a corresponding bipartite Tanner graph.  The girth $g$ is the length of the shortest cycle 
in the Tanner graph.

When decoded over a BEC, the FER performance of the BP decoding is determined by the size of the smallest stopping set  called stopping distance $d_{\rm{stop}}$ (see, for example, \cite{di2002finite}). In turn, a stopping set is defined as a subset of indices of columns in a parity-check matrix, such that  a matrix  constructed from these columns does not have a row of weight one. The asymptotic behavior of a stopping set distribution for ensembles of binary LDPC codes is studied in \cite{orlitsky2005stopping}.
In this paper, we study both the  average performance of the ensembles of random LDPC codes 
and of QC LDPC codes widely used in practical schemes.  

Two ensembles of random regular LDPC codes are studied below.
First we study  the Gallager ensemble  \cite{gallager} of $(J,K)$-regular LDPC codes, where $J$ and $K$ denote the number of ones in each column and  in each row of the code parity-check matrix, respectively.
Codes of this ensemble are determined by  random parity-check matrices $\bldH$, which consist  
of the strips  $\bldH_{i}$  of width $M=r/J$ rows each,  $ i =1,2,\dots, J$. 
All strips are random column permutations of the strip  where the $j$th row 
contains $K$ ones in positions $(j-1)K+1, (j-1)K+2, \ldots, jK$, for $j = 1, 2, \ldots, n/K$. 
 
Next, we study the ensemble of binary $(J,K)$-regular LDPC codes, which   
is a special case of the ensemble described in \cite[Definition 3.15] {richardson2008modern}.
We refer to this ensemble as  the Richardson-Urbanke (RU) ensemble of ($J,K$)-regular LDPC codes.

For  $a\in\{1,2,...\}$ denote by $a^m$ a sequence $(a,a,...,a)$ of $m$ identical 
symbols $a$. 
In order to construct an $r \times n$ parity-check matrix $\bldH$ of  an LDPC code from the  RU ensemble, one does the following:
\begin{itemize}
\item 
construct the sequence $\bs a=(1^J,2^J,...,n^J)$; 
\item
apply a random permutation $\bs b = \pi (\bs a)$ to obtain a sequence $\bs b=(b_1,...,b_N)$, where $N=Kr=Jn$;
\item 
set to one
the entries in the first row of $\bldH$ in columns $b_1,...,b_K$,
the entries in the second row of $\bldH$ in columns $b_{K+1},...,b_{2K}$, etc. The remaining entries of
$\bldH$ are zeros.
\end{itemize}
 
In fact, an LDPC  code from the RU ensemble is $(J,K)$-regular if for a given permutation $\pi$ all elements of subsequences 
$(b_{iK-K+1},...,b_{iK})$ are different for all $i=1,...,r$.  
It is shown in \cite{litsyn2002ensembles} that the fraction of regular codes among the RU LDPC codes is roughly  
\[
\exp \left\{ -\frac{1}{2} (K-1)(J-1)     \right\}
\]
which means that most of the RU codes are irregular.
In what follows, we ignore this fact  and interpret the RU LDPC codes as the $(J,K$)-regular codes, 
and call them ``almost regular''. 

{
Generally, the design rate $R=1-J/K$ is a lower bound on the  actual code rate since 
the rank of randomly constructed parity-check matrix can be smaller than the number of its rows.
However, in our study the best generated almost regular RU codes always have the rate equal to the design rate. 
For this reason, we do not distinguish between the design rate and the actual rate.
 }
    
In order to construct random binary images of nonbinary ($J,K$)-regular LDPC codes, we use the standard two-stage procedure. It consists of labeling a proper binary base parity-check matrix  by random nonzero elements of the extension of the binary Galois field. In our work, we select a  parity-check matrix of a binary LDPC code from the Gallager or the RU ensembles as the  base matrix.  

In what follows, the Gallager ensembles of binary regular LDPC codes and binary images of nonbinary regular LDPC codes are used only for the theoretical analysis, while for the simulations we use almost regular LDPC codes from the RU ensemble. The reason for this choice is that in the simulations, the RU LDPC ensembles outperform the Gallager LDPC codes with the same parameters.      

\subsection{QC LDPC codes}
The QC LDPC codes represent a class of LDPC codes which is very  intensively used in communication standards.  
Rate $R=b/c$  QC LDPC codes are determined  by a $(c-b)\times c$  polynomial parity-check matrix of their parent convolutional code \cite{johannesson2015fundamentals}
\begin{equation}
\bldH(D)=\left(\begin{array}{cccc}
h_{11}(D)&h_{12}(D)&\dots&h_{1c}(D)\\
h_{21}(D)& h_{22}(D)&\dots&h_{2c}(D)\\
\vdots&\vdots&\ddots&\vdots\\
h_{(c-b)1}(D)& h_{(c-b)2}(D)&\dots&h_{(c-b)c}(D)
\end{array}
\right)
\label{polynom_matr}
\end{equation}
where $h_{ij}(D)$ is  either zero or a monomial entry, that is, $h_{ij}(D)\in \{0,D^{w_{ij}}\}$
with $w_{ij}$ being a nonnegative integer, $w_{ij}\le \mu$, and $\mu=\max_{i,j} \{ w_{ij} \}$ is the syndrome memory.

 {
The polynomial matrix (\ref{polynom_matr}) determines an $[Mc,Mb]$  QC LDPC block code using a
set of polynomials  modulo $D^{M}-1$. By tailbiting the parent convolutional code to length $M > \mu$, we obtain
the binary parity-check matrix
\begin{equation}
\small
\arraycolsep=3pt \def\arraystretch{1.2}
\bs H_{\rm TB}=\begin{pmatrix}
\bs H_{0}&\bs H_{1}&\dots&\bs H_{\mu-1}&\bs H_{\mu}&{\bs 0}&\dots&\bs 0\\
{\bs 0}&\bs H_{0}&\bs H_{1}&\dots&\bs H_{\mu-1}&\bs H_{\mu}&\dots&\bs 0\\
\vdots &                 &\ddots               &\vdots&\vdots                     &\vdots                   &\ddots&\\
\bs H_{\mu}& {\bs 0}&\dots&\bs 0&\bs H_{0}&\bs H_{1}&\dots&\bs H_{\mu-1}\\
\vdots&\ddots&\vdots&\vdots&\vdots&\vdots&\vdots&\vdots\\
\bs H_{1}&\dots&\bs H_{\mu}&{\bs 0}&\dots&\bs 0&\dots&\bs H_{0}
\end{pmatrix},
\label{tb}
\end{equation}
of an equivalent (in the sense of column permutation) TB
code  (see \cite[Chapter 2]{johannesson2015fundamentals}),   where
$\bs H_{i}$, $i=0,1,\dotsc,\mu$, are binary $(c-b)\times c$ matrices in the series expansion
\begin{equation*}
	\bs H(D)=\bs H_{0}+\bs H_{1}D+\cdots+\bs H_{\mu}D^{\mu},
\end{equation*}
and $\bs 0$ is the all-zero matrix of size $(c-b)\times c$. 
If each column of $\bldH(D)$ contains $J$  nonzero elements, and each row contains
$K$ nonzero elements, the QC LDPC block code is $(J,K)$-regular. It is irregular otherwise.
}

Another form of the equivalent $[Mc,Mb]$ binary QC LDPC block code can be obtained by replacing
the nonzero monomial elements of $\bldH(D)$ in (\ref{polynom_matr})  by the  powers of the circulant $M\times M$  permutation matrix $\bldP$, whose rows are cyclic shifts by one position to the right of the rows of the identity matrix.

The polynomial parity-check matrix $\bldH(D)$ (\ref{polynom_matr}) can be interpreted as a $(c-b) \times c$ binary base matrix $\bldB$ labeled by monomials, where the entry in $\bldB$ is one if and only if the corresponding entry of $\bldH(D)$ is nonzero, i.e.
\[
\bldB=\bldH(D)|_{D=1}
\]
All three matrices $\bldB$, $\bldH(D)$, and $\bldH$
can be interpreted as bi-adjacency matrices of the corresponding Tanner graphs. 

 
\section{Stopping redundancy and convergence to the ML decoding performance}\label{redundancy}

The idea to improve the performance of iterative decoding of linear codes over a BEC by 
using redundant parity checks was studied, for example, in \cite{Santhi,Yedidia}. This approach was further explored in
\cite{schwartz2006stopping} (for BEC) and in \cite{jens} (for BSC and AWGN). 
The idea of using redundant parity checks was also studied in the context of linear-programming decoding~\cite{feldman}, the reader can refer, for example, to~\cite{pascal}. 

A straightforward method to extend a parity-check matrix of an LDPC code is based on appending a predetermined number of dual codewords to the parity-check matrix. 
In this approach, the BP decoder uses the redundant matrix instead of the original parity-check matrix.
One of the strategies used to extend the parity-check matrix consists of appending dual codewords in the order of their increasing weights  starting with the minimum weight $d_{\rm dual}$.
A problem of  searching for low-weight dual codewords has high computational complexity in general, yet for short LDPC codes it is feasible.
We apply this approach in the sequel, and study the convergence of the FER of BP decoding of LDPC codes determined by their extended parity-check matrices to the FER of the ML decoding (for both BEC and AWGN channels). 

The stopping redundancy is defined as the minimum number of  rows in a parity-check matrix required to ensure that the  stopping distance of the code $d_{\rm{stop}}$ is equal to  the code minimum distance $d_{\min}$. 
For a set of the selected LDPC codes, we compute estimates on the minimum number of the rows required in order to 
ensure removal of stopping sets of a certain size.  
Next, we describe this approach in more detail. By \emph{$\ell$-th stopping redundancy}, $\rho_\ell$, we denote the minimum number of rows in any parity-check matrix of the code, such that all ML-decodable stopping sets of size less than or equal to $\ell$ are removed. In particular, $\rho_r$ is the minimum number of rows in any parity-check matrix of the code, such that there are no ML-decodable stopping sets of size up to $r$ (incl.), i.e. no stopping sets which, if erased, still can be decoded by the ML decoder. Our definition of $\ell$-th stopping redundancy is analogous to its counterpart in \cite{hehn2008permutation}.

However, we stress the difference between the updated definition of the $\ell$-th stopping redundancy for $\ell \geq d$ and its counterpart in \cite{hehn2008permutation}. In fact, the stopping sets of size $\ell \geq d$ that are \emph{not} ML-decodable, are exactly the supports of the codewords.\footnote{We recall that a support  of a codeword  is a stopping set.}
 
In order to calculate the upper bounds on the $\ell$-th stopping redundancy with a method based on \cite{yakimenka2015refined}, we first estimate by sampling $u_i$, the number of ML-decodable stopping sets of size $i$ in a particular parity-check matrix. Then, we use the estimates on $u_i$ ($i=1,2,\dotsc,r$) with the method similar to \cite[Thm.~1,2]{yakimenka2015refined} in order to obtain the approximate upper bounds on the \emph{stopping redundancy hierarchy}, i.e. the stopping redundancies $\rho_1, \rho_2, \dotsc, \rho_r$. 

In Table \ref{t1}, we present  estimates on $\rho_\ell$, $\ell=d_{\min}, d_{\min} +1,d_{\min}+2$, and $\ell=r$, along with $d_{\min}$, $d_{\rm stop}$, $d_{\rm dual}$ and $g$, for a set of selected LDPC codes.
In Section \ref{simulations}, we also present the simulated FER performance of the BP and ML decoding over the BEC for this set  of codes with varying number of redundant rows. The same set of  LDPC codes with varying number of redundant rows in their parity-check matrices is also simulated over the AWGN channel.

The LDPC codes from the following four families were selected:
\begin{itemize}
\item{Random regular LDPC codes from the RU ensemble (rows 2 and 3 in Table~\ref{t1}) }
\item{QC LDPC codes (row 4)}
\item{Binary images of nonbinary regular LDPC codes (row 5)}
\item{Linear codes represented in a ``sparse form'' (row 1)}
\end{itemize}
Two random RU codes were selected by an exhaustive search among 100000 code candidates. As a search criteria, we used the minimum distance and the first spectrum coefficient $A_{d_{\min},n}$.  The QC LDPC code was obtained by optimization of lifting degrees for a constructed base matrix in order to guarantee the best possible minimum distance under a given restriction on the girth value of the code Tanner graph.  For comparison, we simulated the best linear code with the same length and dimension determined by a parity-check matrix with the lowest possible correlation between its rows. Next, we refer to this form of the parity-check matrix  
as a ``sparse form''. Parameters of the selected codes are presented in Table \ref{t1}. Here we use the notations `RU' for random LDPC codes, `L' for the best linear code with parity-check matrix in `sparse form', `NB' for the binary image of nonbinary regular LDPC code and `QC' for QC LDPC code, respectively.

\begin{table}[h]
  \centering
  \caption{
\label{Table_AWGN}
Parameters of studied $[48.24]$  codes 
}\begin{tabular}{|c|c|c|c|c|c|c|c|c|c|}
\hline
Code& $d_{\min}$&$A_{d_{\min},n}$& $d_{{\rm stop}}$&$d_{\rm {dual}}$&$g$ &$J$,$K$& $\rho_{d_{\min}},\rho_{d_{\min}+1},\rho_{d_{\min}+2}$ &$\rho_r$ &Type\\ \hline   
1 & 12 &17296& 4 & 12 & 4 &6,12 & 6240,12151,23468 & 13\,761\,585 &'L' \\  \hline
2 &8 &13 & 4 & 6 & 4 & 6,12& 261,581,1254 & 13\,683\,513 &'RU' \\  \hline
3 & 7& 1& 5 &5 & 4 & 4,8&  83,175,380 & 12\,549\,204 &'RU' \\   \hline
4 &7&8 & 7 & 5 & 6 & 3,6& 58,130,274 & 9\,876\,964 & 'QC' \\  \hline
5 & 8&7 & 4 & 7 & 4 & 3,6& 355,751,1551 & 13\,819\,276 & 'NB' \\  \hline
\end{tabular}
\label{t1}
\end{table}





\section{Upper bounds on ML and BP decoding  error probability for ensembles of LDPC codes }
\label{bounds}
In this section, we analyze  the Gallager ensembles of binary and binary images of nonbinary  ($J,K$)-regular LDPC codes.  By following  the  approach in \cite{bocharova2017performance} we derive estimates on the decoding error probability of the ML and BP decoding by using  precise coefficients of the average weight spectrum  and average stopping set size spectrum, respectively.  Additionally  to the bounds on the performance of the ML decoding obtained in  \cite{bocharova2017performance}, in this paper we derive the improved bounds on the performance of BP decoding for both binary LDPC codes and binary images of nonbinary regular LDPC codes.  

The main idea behind the approach in \cite{bocharova2017performance} is computing the average spectra coefficients recurrently with complexity linear in $n$. The resulting  coefficients are substituted  into the union-type upper bound on the error probability  of the ML decoding over a BEC \cite{berlekamp1980technology}
\begin{equation} \label{gen_spectr}
P_e \le  \sum_{i=d}^n \min \left\{  \binom{n}{i} ,
\sum_{w=d}^i S_w \binom{n-w}{i-w}
 \right\} \epsilon^i(1-\epsilon)^{n-i}
\end{equation}
where $S_{w}$ is the $w$-th weight (stopping set size) spectrum coefficient, $\varepsilon$ is the erasure probability and $d$ denotes the minimum distance (stopping distance). In order to upper-bound the error probability of the ML decoding over an AWGN channel, the average weight spectrum coefficients are substituted
 into the  tangential-sphere bound \cite{poltyrev1994bounds}.     

Consider  the Gallager ensemble of $q$-ary LDPC codes, where $q=2^{m}$, $m \ge 1$ is an integer.  The weight generating function  of $q$-ary sequences of length $n$ satisfying the nonzero part of one $q$-ary parity-check equation is given in \cite{gallager} as
\begin{equation}
\label{eq:2stars}
g(s)=\frac{(1+(q-1)s)^{K}+(q-1)(1-s)^{K}}{q} \; .
\end{equation}
It is easy to derive the weight generating function of $q$-ary sequences of length $K$ and  $q$-ary weight not equal to 1:
\begin{equation}
\label{stopping}
g_{\rm{stop}}(s)=\sum_{w=0,2,3,...,K}\binom{K}{w}(q-1)^{w}s^{w}=(1+(q-1)s)^{K}-K(q-1)s.\;
\end{equation} 
Each $q$-ary symbol can be represented as a binary sequence (image) of length $m$. It is easy to see that different representations of a finite  field of characteristic two  will lead to different generating functions of binary images for the same ensemble of nonbinary LDPC codes.  Following the techniques in \cite{el2004bounds},  we study an average binary weight spectrum for the ensemble of $m$-dimensional binary images. By assuming  
uniform distribution  on the 
$m$-dimensional binary images of  the non-zero $q$-ary symbols, we obtain  the generating function of the average   binary weights of a $q$-ary symbol in the form 
\begin{equation}
\label{eq:3stars}
\phi(s)= \frac{1}{q-1} \sum_{w=1}^m\binom{m}{w}s^{w}=\frac{(1+s)^{m}-1}{q-1} \; .
\end{equation}
The  average binary weight  generating function for one strip  is given by
\[
G(s)= \big( g( \phi(s)) \big)^{M}=\sum_{w=0}^{nm}N_{nm,w}s^{w} \; , 
\] 
where $N_{nm,w}$ denotes the average number of binary sequences $\bs \beta$ of weight $w$ and of length $nm$ satisfying ${\bs\beta}\mathcal{B}_{i}^{\rm T}=\bs 0$. Here,  $\mathcal B_{i}$ denotes  the average  binary image of  $\bldH_{i}$. We obtain the average binary weight enumerator of nonbinary regular LDPC code  as
\begin{equation} {\rm E} \{A_{nm,w} \}=\binom{nm}{w}\big(p(w)\big)^{J}
=\binom{nm}{w}^{1-J}N_{nm,w}^{J},\label{nonbin} \end{equation}
where 
$
p(w)={\binom{nm}{w}} ^{-1}{N_{nm,w}} .
$
By substituting  (\ref{eq:3stars}) into (\ref{stopping}), similarly to (\ref{nonbin}), we obtain the average binary stopping set size spectrum coefficient. 

It is known  that if the generating function is represented as a degree of another 
generating
function it can be easily computed by applying a recurrent procedure. Details of the recurrent procedure for computing coefficients of the average weight spectra can be found in \cite{bocharova2017performance}.  We proceed by computing $N_{nm,w}$ recursively.  


\section{Simulation results}\label{simulations}
\begin{figure}
\begin{center}
\includegraphics[width=120mm]{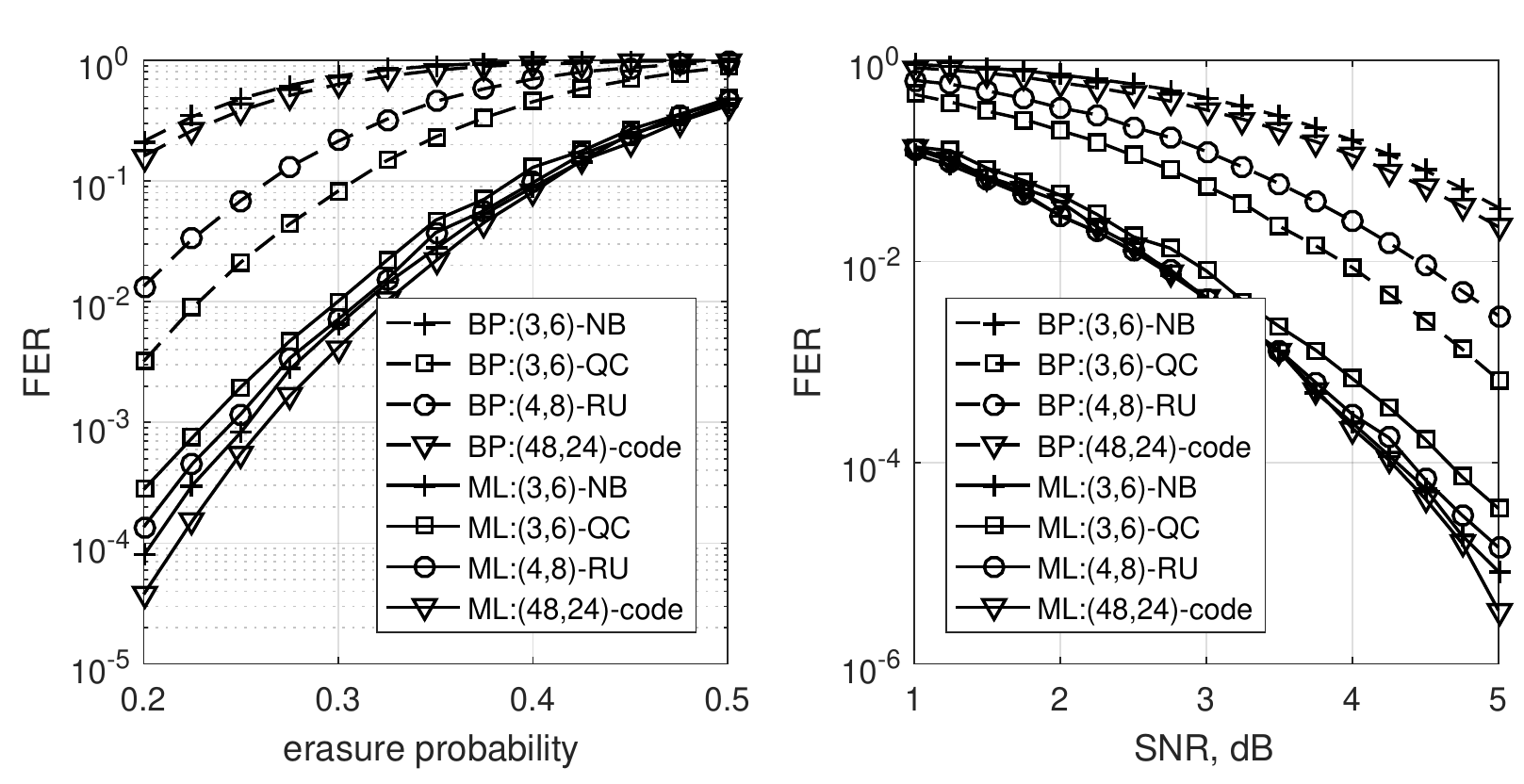}  
\caption{\label{comp_BEC_AWGN}  Comparison of the FER performance of BP and ML decoding over the BEC and the AWGN channel for  LDPC codes of length $n=48$ and rate $R=1/2$
  }
\end{center}
\end{figure}

We simulate the BP and ML decoding over the BEC and AWGN channel for the five LDPC codes whose parameters are presented in Table \ref{t1}. In Fig. \ref{comp_BEC_AWGN}, the FER performance of the BP and ML decoding over the BEC and the AWGN channel is  compared. It is easy to see that the best BP decoding performance both over the BEC and over the AWGN channel (and at the same time the worse ML decoding performance) is  shown by the QC LDPC code with the most sparse parity-check matrix and the largest girth value of its Tanner graph. We remark that the best linear [48,24,12] code determined by a parity-check matrix in a ``sparse form'', as expected, has the best ML decoding performance over the both channels. 
Its BP decoding performance is worse than that of the selected LDPC codes except for the binary image of nonbinary LDPC code.

\begin{figure}
\begin{center}
\includegraphics[width=120mm]{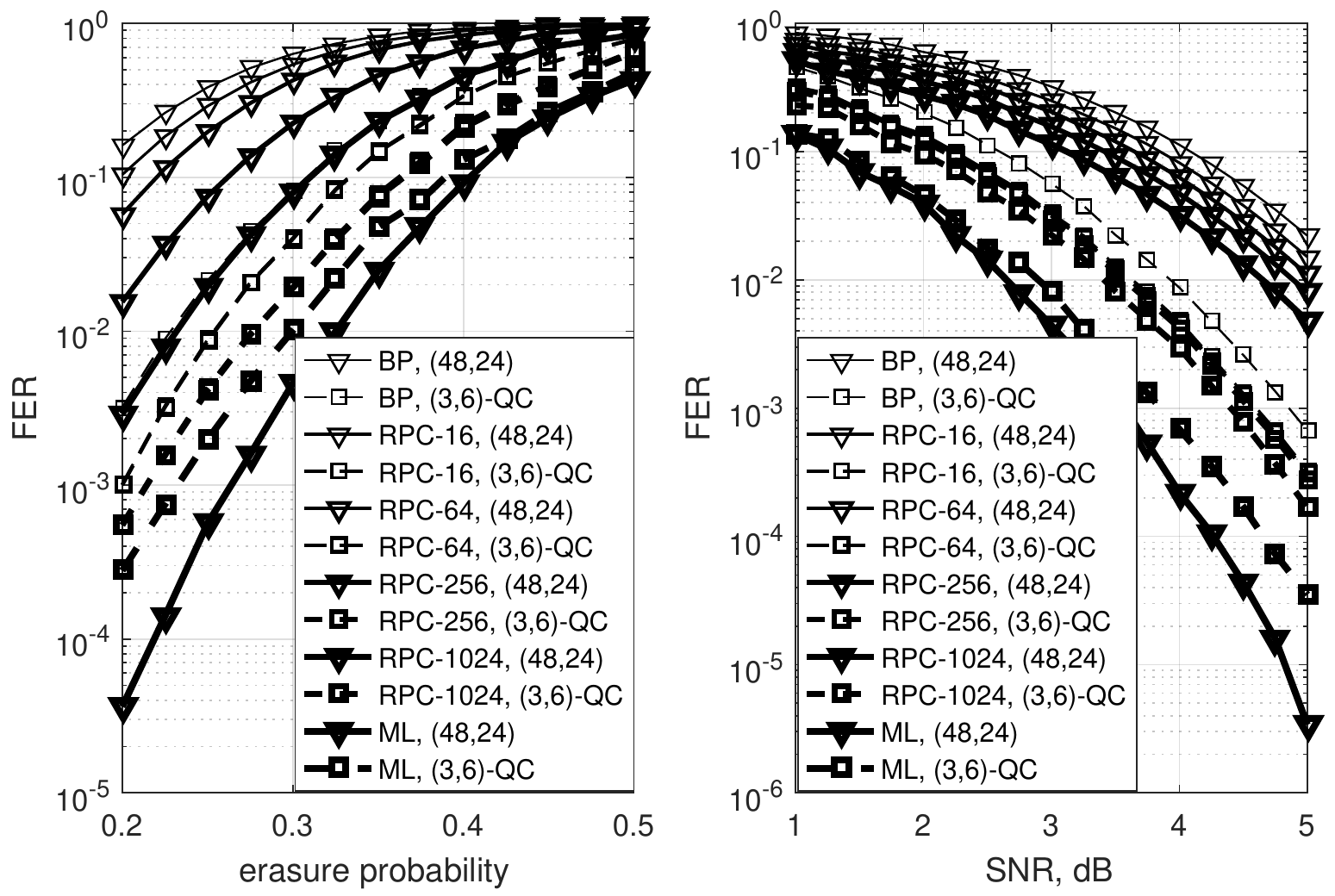}
\caption{\label{rpc_BEC_AWGN} FER performance of RPC  decoding over the BEC  and the AWGN channel for `L' and `QC' codes. 
  }
\end{center}
\end{figure}

Fig. \ref{rpc_BEC_AWGN} shows the BP decoding performance over the BEC and AWGN channel of the codes `QC' and `L' from Table \ref{t1}, when their parity-check matrices are extended.  We call the corresponding decoding technique ``redundant parity check'' (RPC) decoding. The number next to ``RPC'' in Fig. \ref{rpc_BEC_AWGN} indicates the number of redundant rows that was added. The best convergence of the FER performance of the BP decoding over the BEC to that of the ML decoding  is demonstrated by the QC LDPC code, while the best linear code has the slowest convergence of its BP performance to the ML decoding performance. We observe  that the obtained simulation results are consistent with the estimates on the stopping redundancy hierarchy given in Table \ref{t1}.  Surprisingly, similar behavior can also be observed for  the FER performance of RPC decoding  over the AWGN channel. 

%
%

\section{Discussion}\label{discussion}  
\begin{figure}
\begin{center}
\includegraphics[width=90mm]{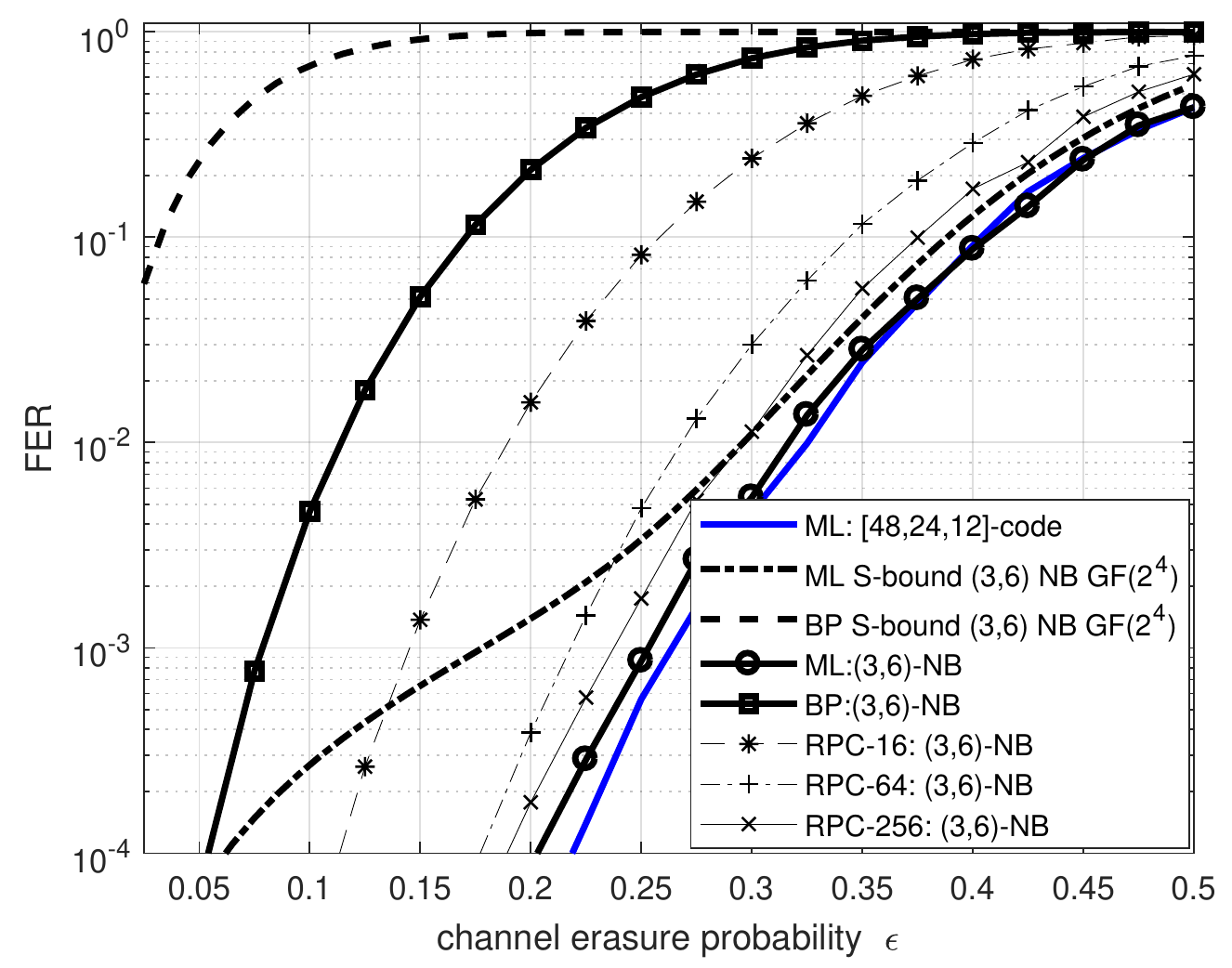}
\caption{\label{NB} Comparison of the  FER performance of BP and RPC decoding over the BEC  with improved union-type bounds  (\ref{gen_spectr}) on the ML and BP decoding performance. 
  }
\end{center}
\end{figure}

In this section, we compare the simulated FER performance of the BP, ML and near-ML (RPC) decoding over the BEC and the AWGN channel with improved bounds on the ML and BP decoding performance. In Fig. \ref{NB}, the FER performance over the BEC for  the binary image of nonbinary $(3,6)$-regular LDPC code over $GF(2^{4})$  (`NB' code in Table \ref{t1}) and the corresponding bounds are shown. 

As it is shown in the presented plots, the ML performance of the `NB' code is rather close to the ML  performance of the `L' code, but the convergence of the FER performance of the RPC decoding to the performance of the ML decoding for the `NB' code is much faster than for the `L' code. 

\begin{figure}
\begin{center}
\includegraphics[width=120mm]{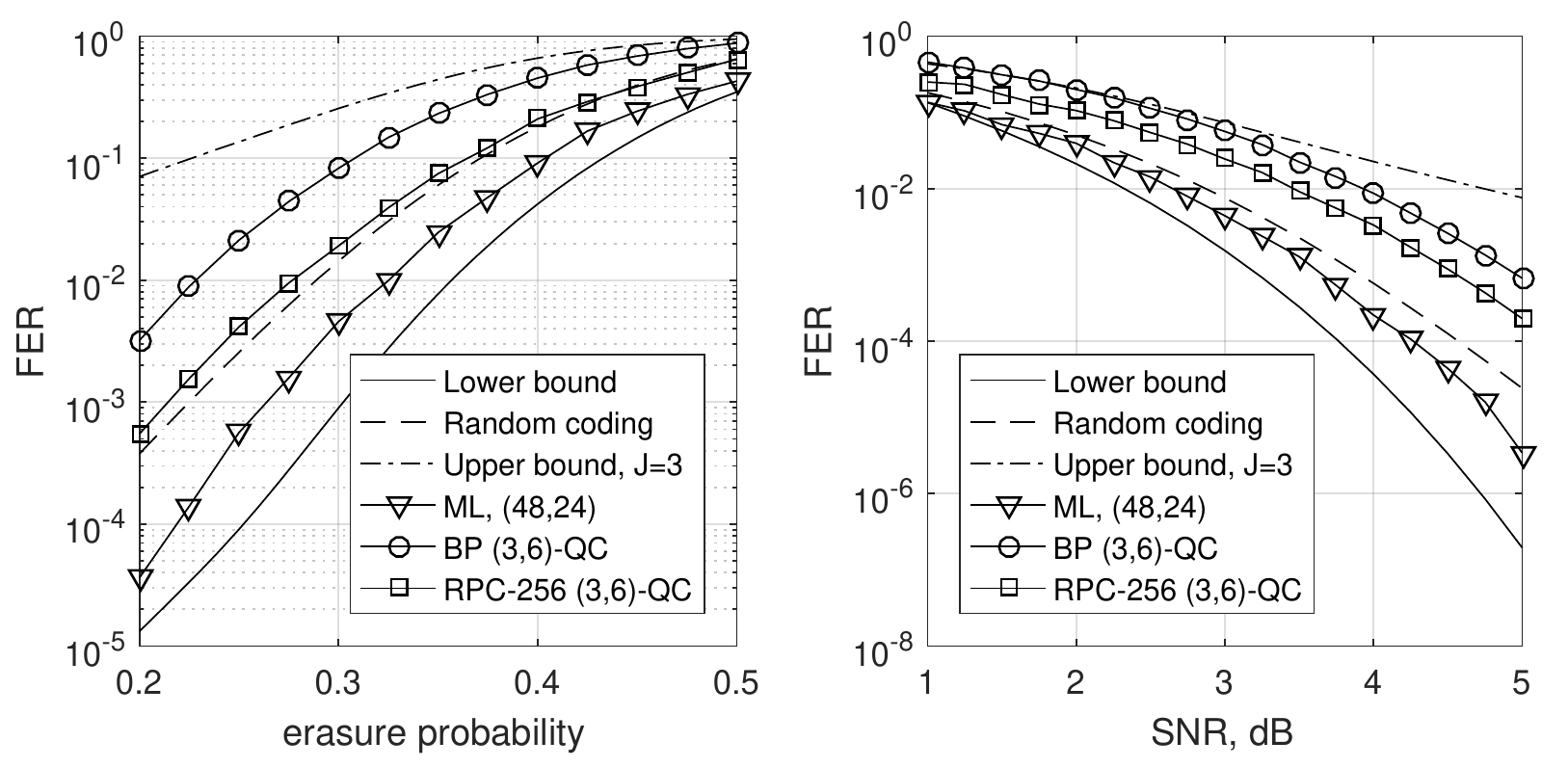}
\caption{\label{Bin_b}Comparison of the  FER performance of BP, ML and RPC decoding with  upper and lower bounds on the ML decoding performance. 
  }
\end{center}
\end{figure}

In Fig. \ref{Bin_b}, the FER performance of the BP, ML and RPC decoding over the BEC and the AWGN channel
is compared to the corresponding upper and lower bounds on the performance of the ML decoding. In particular, for comparison of the performance  over the BEC, we use the improved upper bound (\ref{gen_spectr}) computed for  the precise ensemble average spectrum coefficients for both random linear code and $(3,6)$-regular random binary LDPC code. As a lower bound, we consider the tighten sphere-packing bound in \cite{our_MLbounds}.  For comparison of the performance over the AWGN channel, we show the tangential-sphere upper bound \cite{poltyrev1994bounds} computed with the precise  ensemble average spectrum coefficients for the same two ensembles and the Shannon lower bound \cite{Shannon1959}. 

Based on the presented results, we conclude  the following:
\begin{itemize}
\item{Although it is commonly believed  that the stopping sets influence the BP decoding  performance 
over the BEC only,  the behavior of the analyzed codes over the BEC and the AWGN channel is very similar. In particular, for short codes, the FER performance of the BP decoding over the AWGN channel can be significantly improved by adding redundant rows to the parity-check matrix.}
\item{Convergence of the RPC decoding performance to the ML decoding performance is faster  for those codes which are most suitable for iterative decoding, that is,
codes with large girth of the Tanner graph.}
\item{RPC decoding has a decoding threshold. When a small number of redundant rows is added, the FER performance rapidly improves, but after adding a certain number of redundant rows, the performance improvement becomes practically unjustified due to growing complexity.}    
\item{The FER performance of the RPC decoding achieves the FER performance of the ML decoding over the BEC with exponential (in length) complexity. However, a significant reduction in the FER compared to the FER of BP decoding can be achieved with a   significantly lower complexity than that of the ML decoding.} 
\item{Binary images of nonbinary LDPC codes with RPC decoding demonstrate good FER performance over the BEC. In order to apply RPC decoding to these codes over the AWGN channel it is required to add $q$-ary parity-checks to their parity-check matrices. This method looks promising and is subject of our future research.}      
\end{itemize}

\end{document}